\documentclass{article}
\usepackage{amsthm, amsmath, amsfonts, amssymb}
\theoremstyle{plain}
\newtheorem{theorem}{Theorem}[section]

\newtheorem{corollary}[theorem]{Corollary}

\theoremstyle{definition}

\theoremstyle{remark}
\newtheorem{remark}[theorem]{Remark}

\def\squareforqed{\hbox{\rlap{$\sqcap$}$\sqcup$}}
\def\qed{\ifmmode\squareforqed\else{\unskip\nobreak\hfil
\penalty50\hskip1em\null\nobreak\hfil\squareforqed
\parfillskip=0pt\finalhyphendemerits=0\endgraf}\fi}

\newenvironment{proofof}[1]{\begin{trivlist}%
\item[]{\flushleft\em Proof of #1. }}
{\end{trivlist}}

\def \lket {|}
\def \rket {\rangle}
\def \lbra {\langle}
\def \rbra {|}

\newcommand{\ket}[1]{\lket #1\rket}
\newcommand{\bra}[1]{\lbra #1\rbra}

\newcommand{\fp}{h}
\newcommand{\fpg}{h^G}

\begin{document}
\title{Distributed construction of quantum fingerprints}

\author{Andris Ambainis\footnote{Institute of Mathematics and Computer Science,
University of Latvia, Raina bulv. 29, Riga, LV-1459, Latvia. E-mail: ambainis@lanet.lv.
Part of this research done while at Institute for Advanced Study, Princeton, supported by
NSF Grant CCR-9987845 and the State of New Jersey.}
\and 
Yaoyun Shi\footnote{Department of Electrical Engineering and
Computer Science, University of Michigan, 1301 Beal Avenue,
Ann Arbor, Michigan 48109-2122, USA. E-mail: shiyy@eecs.umich.edu.
Part of this research was done at Institute of Quantum Information, California
Institute of Technology, and was supported by NSF EIA-0086038, NSF 
CCR-0049092, and The Charles Lee Powell Foundation.}}
\maketitle
\begin{abstract}
\noindent
Quantum fingerprints are useful quantum encodings
introduced by Buhrman, Cleve, Watrous, and de Wolf 
({\em Physical Review Letters}, Volume 87, Number 16, Article 167902, 2001)
in obtaining an efficient quantum communication protocol.
We design a protocol for constructing the fingerprint in
a distributed scenario. As an application, this protocol
gives rise to a communication
protocol more efficient than the best known classical
protocol for a communication problem.
\end{abstract}
\section{Introduction}
The fundamental difference between quantum and classical
information has been demonstrated in many aspects of 
quantum information processing. This article concerns
that of {\em communication complexity}, where
two parties, Alice, who holds $x\in\{0, 1\}^n$, 
and Bob, who holds $y\in\{0, 1\}^n$, wish to compute a function
$f(x, y)$.
The minimum amount of information they need to exchange
for all considered inputs is the communication complexity of $f$.
Classical communication complexity has been widely studied since
its introduction by Yao~\cite{Yao:1979:comm}. 
An excellent book on the subject is by Kushilevitz and 
Nisan~\cite{Kushilevitz:1997:book}.
In recent years, many works on quantum communication
complexity, also introduced by Yao~\cite{Yao:1993:circuit},
have shown both the power and the limitations of quantum
communication protocols (e.g. \cite{Buhrman:1998:com, 
Razborov:2002:disj}).

We shall focus on a variant of communication complexity models
called {\em Simultaneous Message Passing} (SMP) model,
where Alice and Bob send a single message to a referee, who will
then determine $f(x, y)$. In an elegant paper\cite{Buhrman:2001:QF},
Buhrman, Cleve, Watrous and de Wolf showed that there is quantum
protocol that uses $O(\log n)$ qubits to compute the \textsc{Equality}
problem, i.e., checking if $x=y$, in this model, while
the best classical protocol requires $\Theta(\sqrt{n})$ bits
\cite{Ambainis:1996:comm, Newman:1996:comm, Babai:1997:comm}. 
This is the first exponential separation between
quantum and classical communication for computing a total function.

At the heart of their protocol is an interesting quantum object called
{\em quantum fingerprint}. For any $z\in\{0, 1\}^n$, the quantum 
fingerprint of $z$ is $\ket{\fp_z} = \frac{1}{\sqrt{n}}
\sum_{i=1}^n (-1)^{z_i}|i\rangle$, where $z_i$ is the $i$th bit
of $z$. The first step in the protocol is to encode inputs 
$z\mapsto z'\in\{0, 1\}^{kn}$ by a linear code that has code
distance $\ge \lambda kn$, where $k$ is an integer and $\lambda>0$.
Then Alice sends $\ket{\fp_{x'}}$ and Bob sends $\ket{\fp_{y'}}$,
upon receiving which the referee will check if $|\langle \fp_{x'} |
\fp_{y'}\rangle|=1$ or bounded away from $1$. This gives the answer
for whether $x=y$.

In this paper we consider the communication complexity of 
constructing $\ket{\fp_{x+y}}$ in the SMP model. That is,
Alice sends a message $\ket{\phi_x}$, and Bob $\ket{\phi_y}$,
to the referee, who will construct a (mixed) state close to 
that of $\ket{\fp_{x+y}}$. Our main result is the following.

\begin{theorem}
\label{thm:main}
Suppose that Alice holds $x\in\{0, 1\}^n$ and Bob
holds $y\in\{0, 1\}^n$. 
For any constant $\epsilon>0$, there is a SMP protocol 
of $O(\sqrt{n}\log n)$ qubits with success
probability $\ge 1-\epsilon$, and if the protocol succeeds
the mixed state $\rho$ of the referee satisfies
$\langle \fp_{x+y} | \rho | \fp_{x+y}\rangle \ge 1- O(1/\sqrt{n})$.
\end{theorem}

The Equality problem can be generalized to the co-linear checking
problem $\textrm{Lin}_k$, where $k$ players, who hold $n$ bit strings $x_1$,
$x_2$, $\cdots$, $x_k$, would like to check if $\sum_{i=1}^k x_i=0$,
by sending a single message from each of them to a referee.

Several variants of co-linearity testing are possible.
We can require $\sum_{i=1}^k x_i=0$ with addition in $Z_2^n$ (bitwise
XOR of $n$-bit strings), in $Z_N$ (modulo some integer $N>2^n$)
or, more generally, with addition in some group $G$.
If addition is in $Z_2^n$, there is 
a classical protocol for $\textrm{Lin}_4$
that communicates $\Theta(n^{3/4})$ bits and for $Lin_k$
that communicates $\Theta(n^{(k-1)/k})$ bits.
(The protocol is a simple generalization of 
\textsc{Equality} protocol in \cite{Ambainis:1996:comm, Newman:1996:comm, Babai:1997:comm}.)
We do not know if this protocol is optimal.
For addition in $Z_N$ or arbitrary groups, no classical
protocol with $o(n)$ communication is known.
(The simple generalization of \textsc{Equality} protocol no longer works.)
In contrast,

\begin{theorem}
\label{thm:colinear}
For any Abelian group $G$, $|G|\leq 2^n$ there is a 
$O(\sqrt{n}\log n)$ qubit quantum protocol 
for $\textrm{Lin}_4$ with addition in group $G$.
\end{theorem}

This improves over the best known classical protocols both
for $Z_2^n$ and for arbitrary group $G$. The proof for 
$Z_2^n$ is a straightforward combination of \cite{Buhrman:2001:QF}
with our distributed fingerprint construction.
The proof for arbitrary $G$ requires a different
construction of fingerprints.

We can also generalize Theorems \ref{thm:main} and \ref{thm:colinear}
to larger number of players. If $k$ players would like to construct
a fingerprint for the sum of $k$ inputs, each of them belonging
to one party, it suffices to communicate $O(n^{1-1/k}\log n)$ qubits.
As a consequence, this gives an $O(n^{1-1/k}\log n)$ communication
protocol for $Lin_{2k}$, for any Abelian $G$, $|G|\leq 2^n$. 


\section{Protocol for distributed fingerprinting}
We shall prove Theorem \ref{thm:main} first, then discuss
what the protocol can be used for. 

\begin{proofof}{Theorem \ref{thm:main}}
The protocol is as follows (we do not scale a pure state to be a unit
vector).  Alice sends the state
\[ \sum_{A\subseteq [n], |A|=\sqrt{n}} (-1)^{\sum_{i\in A} x_i} \ket{A} \]
to the referee and Bob sends
\[ \sum_{B\subseteq [n], |B|=\sqrt{n}} (-1)^{\sum_{i\in B} y_i} \ket{B}.\]
This requires communicating 
$\log {n \choose \sqrt{n}}=\Theta (\sqrt{n}\log n)$ qubits.

After receiving the states from Alice and Bob, the referee projects
the state to the subspace spanned by $\ket{A}\ket{B}$
satisfying $|A\cap B|=1$.
If he gets a state not in this subspace, he outputs ``fail''.
The probability of not failing is just the probability
that two random sets of size $\sqrt{n}$ have intersection of size 1.
This probability is at least $1/e$.
Therefore, by repeating the protocol for a constant times,
the probability that all pairs of messages fail can be made arbitrary small.
If the protocol does not fail for a pair of messages, the remaining state is
\[ \ket{\phi}= \mathop{\mathop{\sum_{A, B\subseteq [n],}}_{|A|=|B|=\sqrt{n},}}_{|A\cap B|=1} 
(-1)^{\sum_{i\in A} x_i} (-1)^{\sum_{i\in B} y_i} \ket{A}\ket{B} =
\sum_j (-1)^{x_j+y_j} \ket{j} \ket{\phi_j} \]
where
\[ \ket{\phi_j}=
\mathop{\mathop{\sum_{A', B'\subseteq [n],}}_{|A'|=|B'|=\sqrt{n}-1,}}_{|A'\cap B'|=0, 
j\notin A', j\notin B'} 
(-1)^{\sum_{i\in A'} x_i} (-1)^{\sum_{i\in B'} y_i}\ket{A'}\ket{B'} .\]
Let
\[ \ket{\phi_0}=
\mathop{\mathop{\sum_{A', B'\subseteq [n],}}_{|A'|=|B'|=\sqrt{n}-1,}}_{|A'\cap B'|=0} 
(-1)^{\sum_{i\in A'} x_i} (-1)^{\sum_{i\in B'} y_i}\ket{A'}\ket{B'} .\]
Then, $\lbra \phi_j | \phi_0\rket = 1-O(\frac{1}{\sqrt{n}})$.
Therefore, the inner product between $\ket{\phi}$ and
$\sum_j (-1)^{x_j+y_j} \ket{j}\otimes \ket{\phi_0}$
is $1-O(\frac{1}{\sqrt{n}})$ and tracing out the second
part of $\ket{\phi}$ leaves the first part in a mixed state having
overlap $1-O(\frac{1}{\sqrt{n}})$ with the state 
$\ket{\psi}= \sum_j (-1)^{x_j+y_j} \ket{j}$.
\end{proofof}

We refer to the above protocol by Protocol 1.
\begin{corollary}
Let $\ket{\psi}=\frac{1}{\sqrt{n}}
\sum_{i=1}^n (-1)^{x_i+y_i} \ket{i}$
with Alice holding $x_1, \ldots, x_n$ and
Bob holding $y_1, \ldots, y_n$.
Then, Alice, Bob and referee can generate
a mixed state $\rho$ such that 
$\lbra \psi | \rho | \psi \rket\geq 1-O(\frac{1}{\sqrt{n}})$
by communicating $O(\sqrt{n}\log^2 n)$ qubits from Alice and Bob
to referee.
\end{corollary}

\begin{proof}
We can repeat Protocol 1
$\lceil\frac{1}{2\epsilon}\ln n\rceil$ times. 
Then, the probability that all $\lceil\frac{1}{2\epsilon}\ln n\rceil$
executions fail is
\[ (1-\epsilon)^{\lceil\frac{1}{2\epsilon}\ln n\rceil} \leq 
e^{\frac{\ln n}{2}} = \frac{1}{\sqrt{n}} \]
and, if at least one of them does not fail, we can take the
state $\rho$ output by the first execution which does not
fail and it satisfies 
$\lbra \psi |\rho| \psi \rket\geq 1-O(\frac{1}{\sqrt{n}})$.
\end{proof}

\begin{remark}
The result generalizes to the case with $k$ parties instead of Alice and Bob.
Assume that we have $k$ parties with the $i^{\rm th}$ party holding the input
$x^i=(x^i_1, \ldots, x^i_n)$. 
They want to generate the state
\[ \ket{\psi}= \sum_{j=1}^n (-1)^{\sum_{i=1}^k x^i_j} \ket{j} \]
by sending messages to a referee ($(k+1)^{\rm st}$ party).
Then, we can show that the $k$ parties together with
the referee can generate a state $\rho$ satisfying 
$\bra{\psi}\rho\ket{\psi} = 1- O(\frac{1}{\sqrt[k]{n}})$
by communicating $O(n^{\frac{k-1}{k}}\log^2 n)$ qubits.
To do that, they perform a similar protocol
using subsets of size $n^{\frac{k-1}{k}}$ instead of $n^{\frac{1}{2}}$.
Specifically, the $i^{\rm th}$ party sends the uniform superposition
\[ \sum_{A_i\subseteq [n], |A_i|=n^{1-1/k}} (-1)^{\sum_{j\in A_i} x^i_j} \ket{A_i}. \]
Then with at least a constant probability,
$\bigcap_{i=1}^k A_i$ contains exactly one element, in which
case the protocol is considered successful. 
The proof for Theorem~\ref{thm:main}
can be modified accordingly to show 
that this generalized protocol works for $k$ parties.
\end{remark}

\section{Application: colinearity testing in $Z_2^n$}

We now prove the theorem \ref{thm:colinear} when addition is in $Z_2^n$,
following the approach of \cite{Buhrman:2001:QF}.

Choose a linear error correcting code that maps
$z\in\{0, 1\}^n$ to $z'\in\{0, 1\}^{kn}$,
and the code distance $\ge \lambda kn$, for
some integer $k$ and some real number $\lambda>0$.
Such codes exist \cite{Justesen:1972:codes}. Let $\epsilon$ 
be a small constant. 

Then we run Protocol 1 with success probability 
$\geq 1-\epsilon$ among Players 1 and 2, and the referee
to construct the fingerprint for  $x'_1+x'_2$, and 
run Protocol 1 with success probability 
$\geq 1-\epsilon$ among Players 3, 4, and the referee to construct
the fingerprint for $x'_3+x'_4$. Finally the referee runs
the SWAP test \cite{Buhrman:2001:QF} on the two fingerprints 
to check if $x'_1+x'_2=x'_3+x'_4$.
Then, the probability of making error is $O(\epsilon)$ in addition
to the probability that the SWAP test makes error. 
Therefore, the overall error probability can be made arbitrary small
by a repetition of the protocol.

\section{Application: colinearity testing in arbitrary $G$}

The key difference is that Players 1 and 2 now have to generate a fingerprint
for $x'_1+x'_2$, with addition taken in group $G$. 
This requires a different construction of fingerprints \cite{Ambainis:unpublished}.

Let $G$ be a finite Abelian group. Let $\chi_i(g)$ be the group
characters of $G$. We take 
\[ \ket{\fpg_g}=\sum_{i=1}^m \chi_i(g)\ket{i} \]
as the fingerprints, with $m=O(\log |G|)$ and
$\chi_1, \ldots, \chi_m$ being $m$ different characters of $G$.
It can be shown that

\begin{theorem}
\cite{Ambainis:unpublished}
For $\chi_1$, $\ldots$, $\chi_m$ chosen uniformly at random
among all characters of $G$, 
we have $|\lbra \fpg_g | \fpg_{g'}\rket| \leq \epsilon$ 
for all $g$, $g'$, $g\neq g'$, with high probability.
\end{theorem}

We now show a distributed construction for the new fingerprints.

\begin{theorem}
\label{thm:anygroup}
If Alice and Bob are given group elements $g_1, g_2$, they can
construct $\ket{\fpg_{g_1+g_2}}$ with $O(n^{1/2}\log n)$ bit communication.
\end{theorem}

\begin{proof}
Alice sends
\[ \mathop{\sum_{A\subseteq [m],}}_{|A|=\sqrt{m}} \prod_{i\in A} \chi_{i}(g_1) \ket{A}. \]
Bob sends
\[ \mathop{\sum_{B\subseteq [m],}}_{|B|=\sqrt{m}} \prod_{i\in B} \chi_{i}(g_2) \ket{B} .\]
The referee projects
the state to the subspace spanned by $\ket{A}\ket{B}$
satisfying $|A\cap B|=1$.
If he gets a state not in this subspace, he outputs ``fail''.
Again, he does not fail with a constant probability.
If the protocol does not fail, the remaining state is
\[ \ket{\phi}= \mathop{\mathop{\sum_{A, B\subseteq [n],}}_{|A|=|B|=\sqrt{n},}}_{|A\cap B|=1} 
\prod_{i\in A} \chi_i(g_1) \prod_{i\in B} \chi_{i}(g_2)  \ket{A}\ket{B} .\]
By the multiplicative property of group characters, 
$\chi_{j}(g_1) \chi_{j}(g_2)=\chi_{j}(g_1+g_2)$.
Therefore,
\[ \ket{\phi} = 
\sum_{j\in [m]} \chi_{j}(g_1 + g_2) \ket{j} \ket{\phi_j} \]
where
\[ \ket{\phi_j}=
\mathop{\mathop{\sum_{A', B'\subseteq [n],}}_{|A'|=|B'|=\sqrt{n}-1,}}_{|A'\cap B'|=0, 
j\notin A', j\notin B'} 
{\prod_{i\in A'} \chi_{i}(g_1) \prod_{i\in B'} \chi_{i}(g_2) \ket{A'}\ket{B'}} .\]
The rest of proof is similar.
We use $O(m^{1/2}\log m)$ qubits of communication and, since
$m=O(n)$, the theorem follows.
\end{proof}

The protocol for $Lin_4$ ($Lin_k$) for arbitrary Abelian $G$
is the same as before, with the construction of theorem \ref{thm:anygroup} instead of 
theorem \ref{thm:main}.

\section{Conclusion}

We have shown how to construct a quantum fingerprint with $O(\sqrt{n}\log n)$
communication in two-party case and with $O(n^{1-\frac{1}{k}}\log n)$
communication in $k$-party case. We conjecture that $\Omega(\sqrt{n})$
is a lower bound for two-party case and $\Omega(n^{1-\frac{1}{k}})$
is a lower bound for $k$-party case. However, showing a lower bound better than $\Omega(\log n)$
is an open problem. 

Another open problem is: can we construct quantum fingerprints with $o(n)$
communication if we require our protocol to be {\em exact}?
That, the state output by the protocol has to be exactly the
fingerprint and not just a state close to fingerprint.
For applications, a state close to fingerprint suffices
because it can only add an $O(\frac{1}{\sqrt{N}})$ term to
the error probability. However, it would be of theoretical interest 
to know if an exact protocol is possible. Also, understanding
exact protocols could be the first step toward proving that our protocol
is optimal or nearly-optimal among all protocols.

We note that, in other settings, exact quantum algorithms can be 
more difficult than quantum algorithms with small error. For example, 
Grover's algorithm \cite{Grover:1996:Search} can search with $O(\sqrt{n})$ 
queries but any zero-error quantum search algorithm requires $n$
queries \cite{Beals:2001:QLB}.

\end{document}